\documentclass[12pt]{article}
\def\la{\mathrel{\mathpalette\fun <}}

\def\fun#1#2{\lower3.6pt\vbox{\baselineskip0pt\lineskip.9pt
\ialign{$\mathsurround=0pt#1\hfil##\hfil$\crcr#2\crcr\sim\crcr}}}

\title{On the charge of the photon}
\author{L.B. Okun \\
ITEP, Moscow, 117218, Russia}
\date{}
\begin{document}
\maketitle

\begin{abstract}

The papers setting upper bounds on the value of electric charge of
the photon are briefly reviewed. The theoretical framework of
these bounds is shown to be incomplete. Hence the bounds seem to
be unreliable.

\end{abstract}

There exists about a dozen of papers questioning the neutrality of
photon and setting an upper limit on its charge. The title of the
first paper \cite{1} was ``Experimental limit on the ``charge'' of
the photon''. Please note that the word ``charge'' was in
quotation marks, which, I suspect, reflected the scepticism of the
authors. They searched for the effect of electric field $V = 20$
kv on the energy of 14.4 KeV photons emitted by Fe$^{57}$ and
investigated with the use of recoilless resonance scattering.
Neither an energy shift nor a line broadening has been observed
allowing the authors to set the limit $e_\gamma/e < 10^{-15}$.

The authors of experiment \cite{2} were searching for small charge
($\la 0.1 e$) of a Millikan type spheroid suspended magnetically
and exposed to electric field. In this experiment the light used
in the suspension system (photons of average energy 2 eV) for
duration of at least 8 hours would increase the charge of spheroid
if photons were charged. Non-observation of any effect led the
authors to an estimate $e_\gamma/e < 10^{-16}$. This result was a
by-product of experiment looking for free quarks and was briefly
described in the last paragraph of the article \cite{2}.

The latest limit $e_\gamma/e < 5 \cdot 10^{-30}$  given by
``Review of Particle Physics'' \cite{3} is based on the idea of
the absence  of anomalous spread of photons arrival time, applied
to the radio pulses from millisecond pulsar PSR/1937+21. The idea
was put forward in 1988 by Cocconi \cite{4}, who obtained
$e_\gamma/e < 2 \cdot 10^{-32}$. In 1994 Raffelt \cite{5} noted
that Cocconi had not taken into account the standard dispertion
effect in the interstellar plasma. By including this effect
Raffelt \cite{5} concluded with $e_\gamma/e < 5 \cdot 10^{-30}$.

A somewhat weaker constraint was derived in 1992 by Cocconi
\cite{6} from the angular spread of photons propagating from
distant compact extragalactic sources: $e_\gamma/e < 2 \cdot
10^{-28}$. In 2004 Kobyshev and Popov \cite{7} suggested that in
this way the limit  could be improved to $e_\gamma/e < 3 \cdot
10^{-33}$.

Limits on $e_\gamma/e$ based on isotropy of the cosmic microwave
background radiation and some additional assumptions were proposed
in 1994 by Sivaram \cite{8} and in 2005 by Caprini et al.
\cite{9}.

The recent laboratory measurement of the deflection of a laser
beam in a modulated magnetic field has allowed to Semertzidis,
Danby, and Lazarus \cite{10} to produce an upper limit $e_\gamma/e
< 8.5 \cdot 10^{-17}$, which is at the level of experiments
\cite{1} and \cite{2}.

With such a diversity of phenomenological approaches all articles
listed above have one common feature: they ignore the basic
changes in quantum electrodynamics which are needed to include the
non-vanishing charge of the photon.

First of all, if photon is charged and principles of Quantum Field
Theory are not violated, then there should exist at least two
kinds of photons with opposite signs of charges. A minimal
Yang-Mills type theory requires also a neutral photon and
SU(2)-symmetry. This calls for SU(2)-multiplets of electrons,
muons, $\tau$-leptons, as well as quarks. Such extra degeneracy
seems to be in contradiction with totality of data on atomic,
nuclear and particle physics.

Note that the minuscule value of the photon electric charge does
not soften the problem of degeneracy.

I have asked one of the authors of refs. \cite{1} - \cite{10},
whether he thinks that a selfconsistent  scheme with a charged
photon exists or could be invented. His answer was: ``I have never
heard of a consistent theory with a charged photon. However,
theory aside, it is important to test our preconceptions, I
suppose''.

I replied: ``But you are testing it in a framework of a
``theory''.''

The point is that the interaction of a charged photon with
magnetic or electric field is described by a vertex with three
photonic lines. If there exists only one kind of photons, say,
with charge $+e_\gamma$, but not $-e_\gamma$, then such a vertex
would violate charge conservation. Moreover the static fields
containing many virtual photons would be macroscopically charged.

Note that if charge is conserved, then in the case of a single
charged photon the vertex with three photons is forbidden and
formulas used to calculate the deflection of photons in magnetic
field are invalid.

Note also that emission of each new photon by an electron would
change the charge of the latter. Hence an infinite number of
different kinds of electrons are needed.

The situation is not so desperate in the case of a triplet of
photons. But it calls for a more comprehensive analysis.

I am grateful to A. Buras, V. Fadin, M.~Krawczyk, and P.~Zerwas
for drawing my attention to the putative charge of the photon.

\end{document}